# SOCIAL NETWORK FOR SMART DEVICES USING EMBEDDED ETHERNET

Ketul Sheth[1], Shreya Shah[2], Darshan Shah[3], Anuja Odhekar[4]

[1]Department of Electronics and Telecommunication Engineering, University of Mumbai, Dwarkadas J. Sanghvi College of Engineering, Mumbai, India
ketulsheth2@gmail.com
[2]Department of Electronics and Telecommunication Engineering, University of Mumbai, Dwarkadas J. Sanghvi College of Engineering, Mumbai, India
shreyashah191291@gmail.com
[3]Department of Electronics and Telecommunication Engineering, University of Mumbai, Dwarkadas J. Sanghvi College of Engineering, Mumbai, India
darshanshah16@gmail.com
[4]Department of Electronics and Telecommunication Engineering, University of Mumbai, Dwarkadas J. Sanghvi College of Engineering, Mumbai, India
anuja.odhekar@djsce.ac.in

## ABSTRACT

*Embedded Ethernet is nothing but a microcontroller which is able to communicate with the network. A design of AVR controller-based embedded Ethernet interface is presented. In the design, an existing SPI serial device can be converted into a network interface peripheral to obtain compatibility with the network. By typing the IP-address of LAN on the web browser, the user gets a web page on screen; this page contains all the information about the status of the devices. The user can also control the devices interfaced to the web server by pressing buttons provided in the web page. This creates a network for easy communication amongst the devices.*

## 1. INTRODUCTION

Microprocessors are ubiquitous and the Internet is pervasive. Then it is no surprise that many developers are connecting their microprocessors to the Internet. In theory, future is going to be a brave new world where it will be possible to program a VCR at home from a web browser. Left the coffee pot on? Dial up the internet on a Web-enabled device and turn it off [16].

Computer communication systems and especially the Internet are playing an important role in the daily life. Using this knowledge many applications are imaginable. Home automation, utility meters, appliances, security systems, card readers, and building controls, which can be easily, controlled using either special front-end software or a standard internet browser client from anywhere around the world [1]. Web access functionality is embedded in a device to enable low cost widely accessible and enhanced user interface functions for the device. A web server in the device provides access to the user interface functions for the device through a device web page [24]. A web server can be embedded into any appliance and connected to the Internet so the

                                                    41

appliance can be monitored and controlled from remote places through the browser in a desktop. Temperature, Pressure, displacement, motion and sound are the most often measured environmental quantities [3-5].

Among these environmental quantities, temperature is the most often measured parameter in industries. For example, some processes work only within a narrow range of temperatures; certain chemical reactions, biological processes, and even electronic circuits perform best within limited temperature ranges. So, it is necessary to measure the temperature and control if it exceeds some certain limit to avoid any misbehavior of the systems. To accurately control process temperature without Extensive operator involvement a temperature control system relies upon a controller, which accepts a temperature sensor [1].

An AVR controller based embedded Ethernet interface system is designed. In the system, the introduced microcontroller ATMEGA32 can communicate with serial data acquisition equipments at the terminal through SPI interface and can transmit data to remote host computer through Ethernet interface [2]. Compared with the system that a host is connected to many serial devices, the task of host is only to complete a single Ethernet communication and its load is lower [15].

## 2. SYSTEM DESIGN

Many embedded systems have substantially different designs according to their functions and utilities. In the design, structured modular design methods adopted and the system is mainly composed of SPI communication module, controller module and Ethernet interface module, as shown in Fig. 1 [2]. The figure shows that the typical architecture of an embedded Ethernet monitor and control using web browser architecture. Here all the devices are connected to the processor and the ADC converters are used to convert analog data into digital data. Serial communication is done in between microcontroller and Ethernet controller ENC28J60. It is connected to LAN cable through RJ 45 registered jack and the whole device is connected to remote PC through internet.

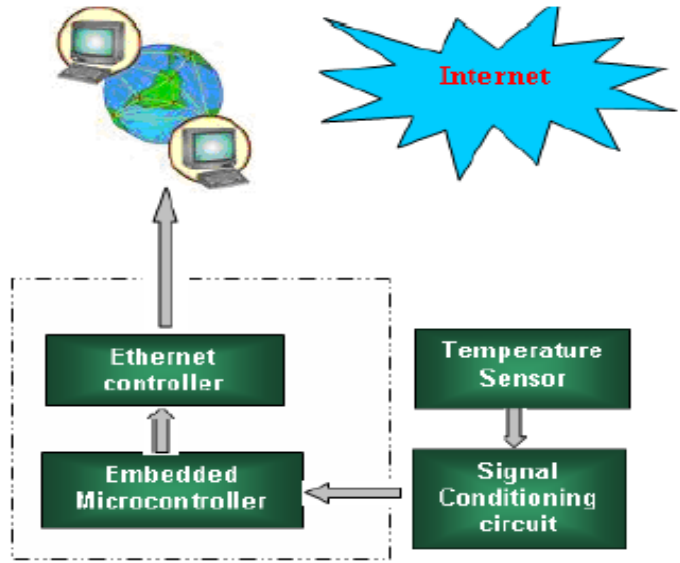

Figure 1.  Basic Block Diagram

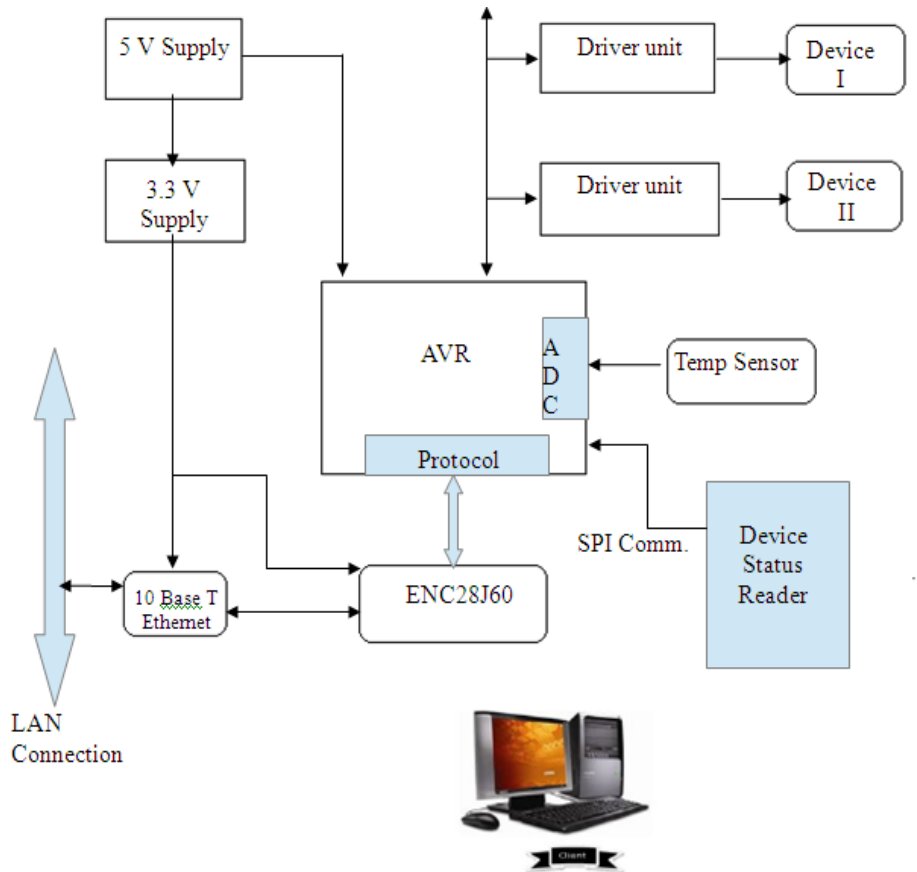

Figure 2. System Architecture

## 2.1. ATmega32 Processor

The ATmega32 provides the following features: 32Kbytes of In-System Programmable Flash Program memory with Read-While-Write capabilities, 1024bytes EEPROM, 2Kbyte SRAM, 32 general purpose I/O lines, 32 general purpose working registers, a JTAG interface for Boundary scan, On-chip Debugging support and programming, three flexible Timer/Counters with compare modes, Internal and External Interrupts, a serial programmable USART, a byte oriented two-wire Serial Interface, an 8-channel, 10-bit ADC with optional differential input stage with programmable gain, a programmable Watchdog Timer with Internal Oscillator, an SPI serial port, and six software selectable power saving modes.

The On chip ISP Flash allows the program memory to be reprogrammed in-system through an SPI serial interface, by a conventional nonvolatile memory programmer, or by an On-chip Boot program running on the AVR core. By combining an 8-bit RISC CPU with In-System Self-Programmable Flash on a monolithic chip, the Atmel ATmega32 is a powerful microcontroller that provides a highly-flexible and cost-effective solution to many embedded control applications. The Atmel AVR ATmega32 is supported with a full suite of program and system development tools including: C compilers, macro assemblers, program debugger/simulators, in-circuit emulators, and evaluation kits [28].

## 2.2. Ethernet Interface

ENC28J60 Ethernet Module is based on Microchip's ENC28J60 stand-alone 10Base-T Ethernet controller. The ENC28J60 is a stand-alone Ethernet controller with an industry standard Serial Peripheral Interface (SPI). It is designed to serve as an Ethernet network interface for any controller equipped with SPI. The ENC28J60 meets all of the IEEE 802.3 specifications. It also provides an internal DMA module for fast data throughput and hardware assisted checksum calculation, which is used in various network protocols. Communication with the host controller is implemented via an interrupt pin, if required, and the SPI, with clock rates of up to 20 MHz. It has dedicated pins for LED link and network activity indication. The module includes RJ45 Ethernet connector with built-in magnetic and link indicator LEDs. It is an ideal device for applications involving Home/Office Automation, Remote Diagnostics systems, Industrial equipments and Security Systems [29].

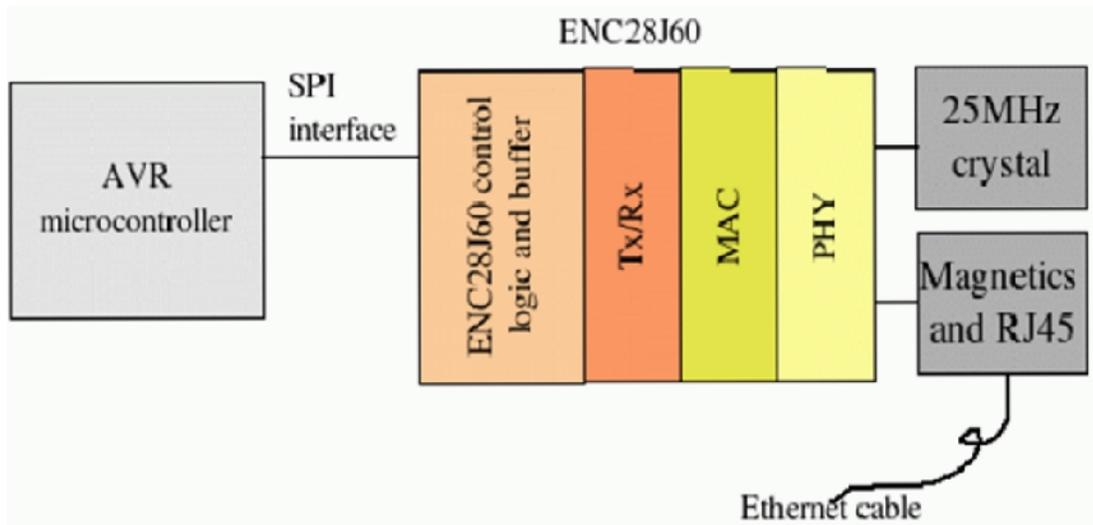

Figure 3. ENC28J60 Network Layers

## 2.3. System Software Design

In order to transmit the data from SPI serial to Ethernet, two system tasks are established. One is to receive front-end data through SPI interface and the other is to transmit data to the Ethernet [21-22].

**SPI receiving task**: For the case that the data are transmit to Ethernet, the data which have been arrived at SPI port are stored into SPI sending buffer and packaged according to TCP/IP protocol and then are added IP and UDP message head on the condition that the PC with SPI interface is set to SPI slave mode and the SPI interface is enabled. At last, the converted data are sent to the host through the corresponding UDP port.

**Ethernet receiving task:** In the Ethernet task, in order to receive the data from Ethernet in the system, the local IP address and subnet mask must be set firstly, and the appropriate UDP port is opened to monitor whether there are data in UDP port. As UDP packet, the data which have been reached the UDP port, are analyzed according to TCP/IP protocol and then stored into SPI receiving buffer. At last, the analyzed data are sent to the SPI serial device through SPI interface driver [7][11].

## 3. SYSTEM TESTING

Embedded Ethernet circuit board is shown in the Fig 4. The circuit board consist of Atmel AVR processor, Ethernet interface module and rj45 registered jack. The devices which are to be controlled are connected to the circuit board.

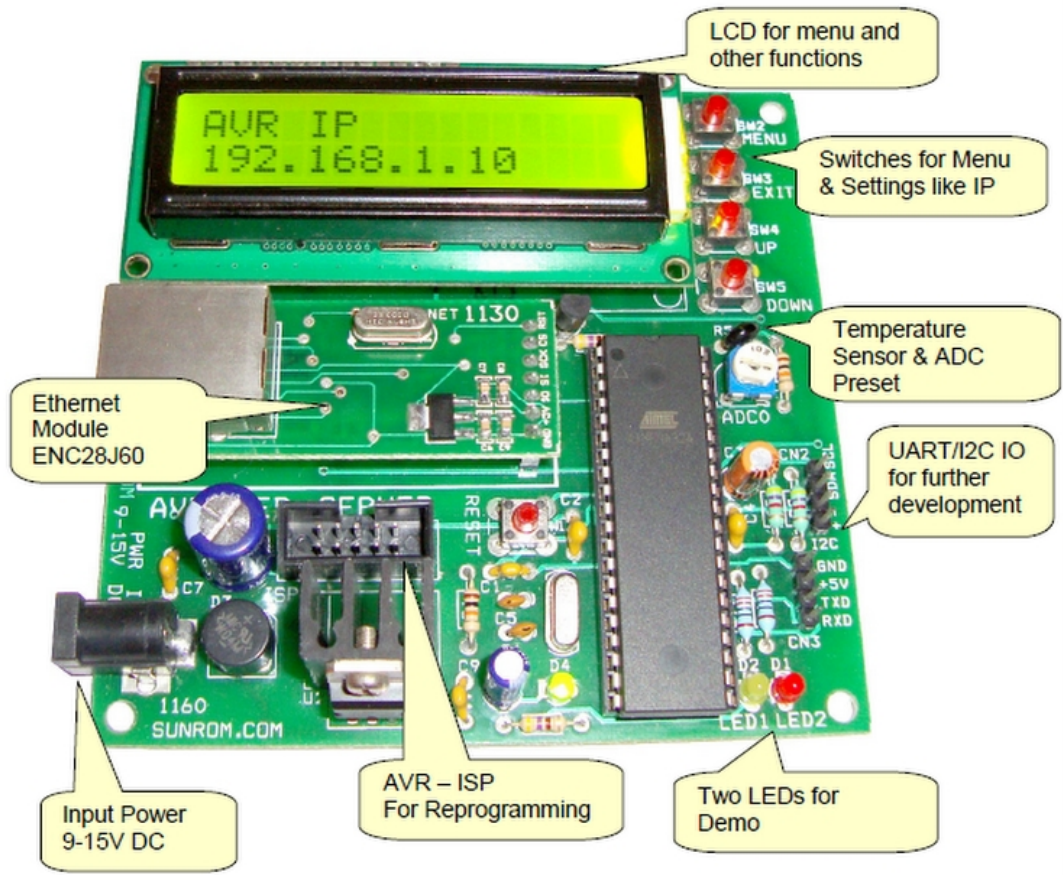

Figure 4. AVR Web Development Board

This board is an embedded AVR Web server. It is based on ATMEGA32 and ENC28J60 chipset. It can serve webpage as well as can be controlled from a PC Application. Temperature sensor can be interfaced to the board, ADC Preset and two LEDs are present to quick start using the board. The reason for using ATmega32 as an 8-bit Microcontroller in our application is because of its high throughput and SPI (Serial Peripheral Interface) interfacing. The SPI interfacing provides serial interface with the device and hence lowering the number of address and data lines used [1].

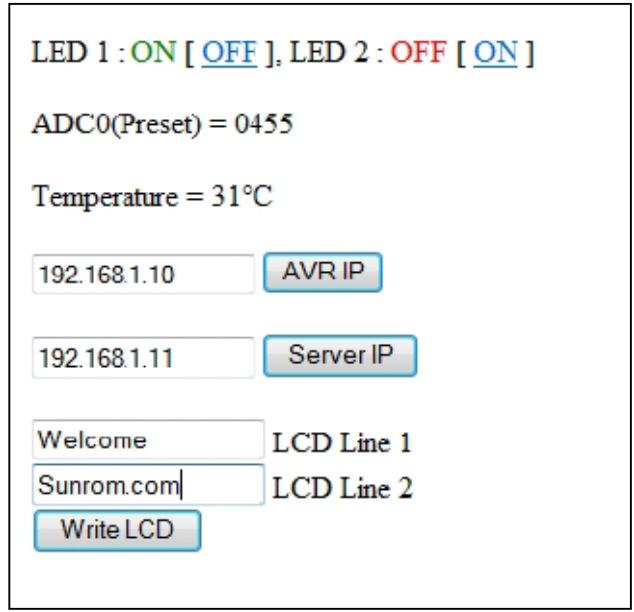

Figure 5. Prototype of web page showing status of devices

Fig. 5 shows the simple web page designed using HTML language. It is requested by the client to server. Then the internet processes these request and server response for client request with web page. Now the Client can know the status of sensors and devices connected to the server and can control the devices via its own browser from remote location. The status of the connected device is shown on the web page by typing the IP address of the server board. We can view the status of the device as shown and change the status by clicking on it and update the system. The status of the temperature of the room will be observed for every hour from the interfaced temperature sensor by refreshing the results. Hence, results show that the client can access the entire set of the connected devices from any remote place via its own local browser. The single AVR board acts as data acquisition and control system and as web server, so the system is compact with less complexity [7].

## 4. SOFTWARE PACKAGES

In this section we will discuss in brief about the software packages used in the project. The software packages used in the project are:

- WinAVR
- AVR Studio

**WinAVR**: WinAVR is a suite of executable, open source software development tools for the Atmel AVR series of RISC microprocessors hosted on the Windows platform. It includes the GNU GCC compiler for C and C++. It provides libraries for various inbuilt functions. WinAVR contains all the tools for developing of an AVR application. This includes Programmers Notepad (editor), avr-gcc (compiler), avrdude (programmer), avr-gdb (debugger) [33].

**AVR Studio:** AVR Studio is the new professional Integrated Development Environment (IDE) for writing and debugging AVR applications in Windows 9x/NT/2000/XP environments. It provides a C compiler, assembler and Simulator [34].

## 5. IMPLEMENTATION

### 5.1 Hardware Implementation

The hardware for the project consists of an AVR Web Development board. The central processor unit is the 8-bit ATMEGA32 microcontroller. It has an onboard Stand-Alone Ethernet controller, ENC28J60 with SPI interface which supports one 10Base-T port, with automatic polarity detection and correction [29]. It is interfaced to an RJ-45 Jack to which the Ethernet cable can be connected. Through this interface our controller can communicate with other devices connected on the network including PCs.

The development board also provides an onboard temperature sensor, ADC preset, two status LEDs, push switches and an LCD module. The LCD display can be used to show the current status of the devices connected to the server and to display the IP address of the server. Using the push switches provided onboard the IP address can be varied. The board can be powered using a 12V 1Amp DC adapter.

The microcontroller can be programmed through the onboard ISP programmer. The required firmware for a microcontroller to act as a web server i.e. the TCP/IP stack is loaded into the microcontroller. To enable the PCs and internet enabled mobile devices to check and control the status of the devices, the development board is connected to the network port of the router. Thus any device in the same network can connect to the web server through the local internet. Also by acquiring DNS from the Internet Service Provider, specifying proper gateway and hosting the webpage, this data can be made available on the Internet [22].

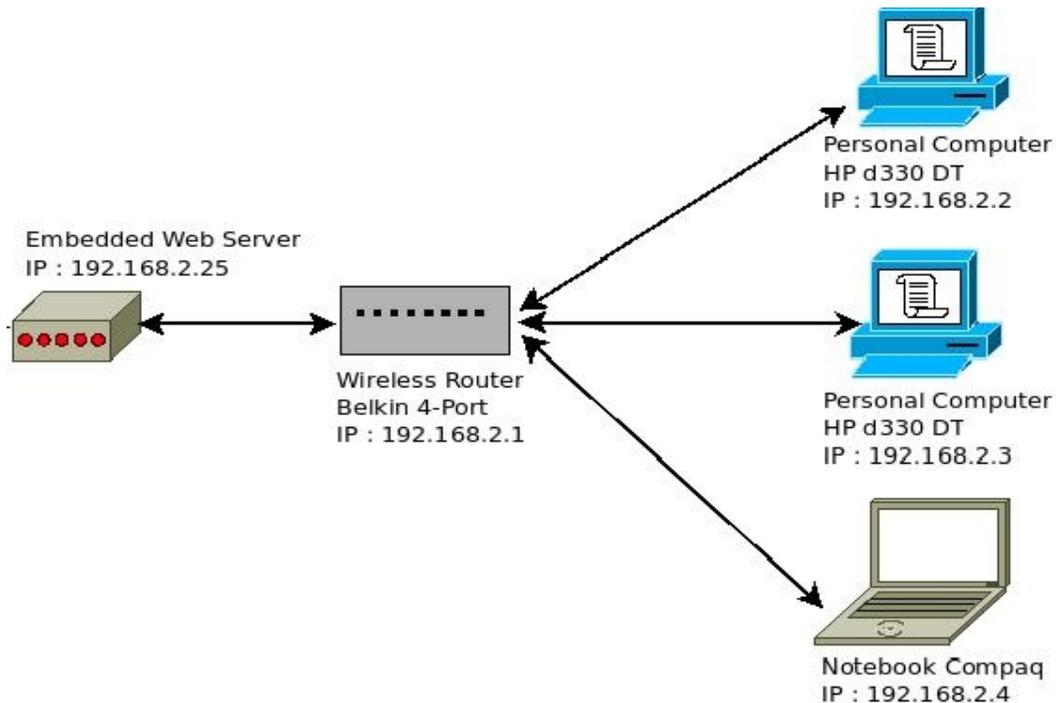

Figure 6. Hardware Implementation

## 5.2 Software Implementation

The header files are written by declaring the necessary Macros and Function into a file ending with .h extension. The real definition of the function appears in the C program with the same file name as header file but ending with .C extension [25]. Most of the header files are obtained under GNU general public license and are rewritten for our convenience. The main header file is where the infinite loop is implemented. It also has some function defined as Server Process which is called by main loop.

All the codes are written on AVR Studio. Each of the function was independently tested with the set of inputs and was simulated by inbuilt Simulator of AVR Studio. After integration of all the functions a sample input was provided to check the output. Also in between code LCD display messages were provided so as to perform debugging through LCD display. When the whole software code is written in C in AVR Studio environment, the output hex file is transferred to the Microcontroller with the help of ISP provided onboard.

Now plug one of the RJ 45 connector of the cross over cable to the RJ 45 jack of AVR web server and the other end connector to the RJ 45 Jack of the router. Manually configure the Internet Connection to a static IP address. Use the Ping command in the client computer to test the system. This test method serves to determine that the device is detected in the network. Notebook client send the ping command to embedded web server device. The web server should respond to the command. Using the static IP address of the server in the client's web browser, the webpage showing the status of the devices can be viewed.

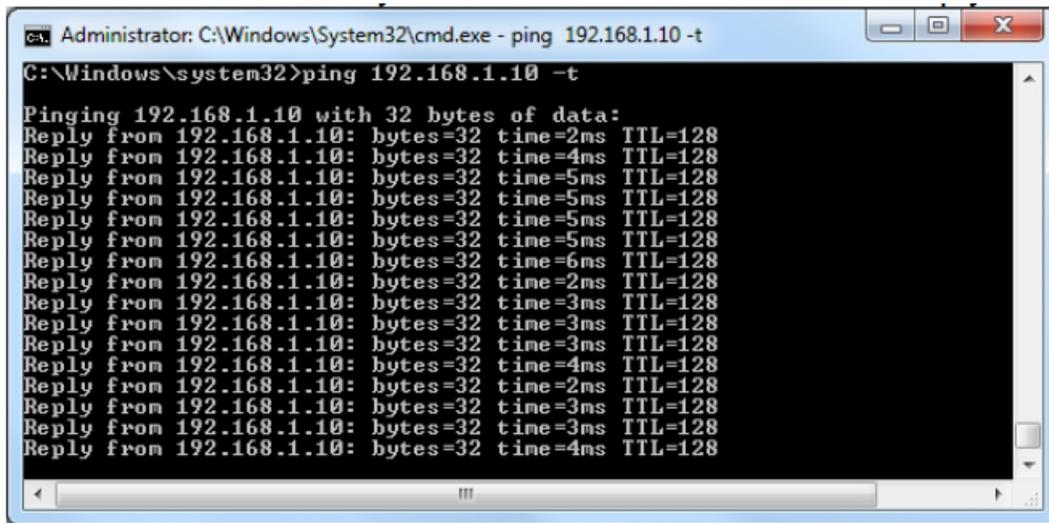

Figure 7. Ping Command Results

**Programming process:**

Step 1. Create an html page to run the microcontroller through. Import it in the code as a string.
Step 2. Set IP, DNS, Gateway addresses and Subnet mask obtained from your internet provider.
Step 3. Initialize the SPI module of the Atmega32 microcontroller.
Step 4. Initialize the Serial Ethernet module chip ENC28J60.
Step 5. Write the code within the Spi_Ethernet_user TCP function that will, after receiving command via web browser, turn on/off the devices connected to the PORT.

Step 6. Read received data in an endless loop. After the web browser "GET" request is received, sent from your computer to the control system IP address, the microcontroller will respond with a web page stored in its memory. This page will then be automatically displayed on the computer screen by the browser.

The advantage of using the Hypertext Transfer Protocol (HTTP) server in the embedded system is; you don't have to develop a special client application to communicate with your embedded system [20]. All you need is to use any standard browser that comes with your personal computer operating system or gadget to talk to your embedded system. The HTTP server uses a simple text called Hypertext Markup Language (HTML) to interact with the browser (client application) through the TCP/IP protocol [19][20].

The HTTP server work by listening to any request from the client (browser) for any HTTP "**GET**" or "**POST**" request through the TCP/IP port **80** (standard HTTP server port). Once the client sends this request to the HTTP server, then the HTTP server will response to this client request by sending the HTTP response header (**HTTP/1.0  200  OK** and **Content-Type: text/html**) follow by the blank line and the HTML text to the client, after transfer all the HTML text to the client; the HTTP server will automatically disconnect the established connection with the client [20]. The following is the example of the client request and the HTML text response transmitted by the embedded HTTP server [30]:

**Client Request:**

GET / HTTP/1.1

Host: 192.168.2.101

User-Agent: Mozilla/5.0 (Windows; U; Windows NT 5.1; en-US; rv:1.9.2.3) Gecko/20

100401 Firefox/3.6.3

Accept: text/html,application/xhtml+xml,application/xml;q=0.9,*/*;q=0.8

Accept-Language: en-us,en;q=0.5

Accept-Encoding: gzip,deflate

Accept-Charset: ISO-8859-1,utf-8;q=0.7,*;q=0.7

Keep-Alive: 115

Connection: keep-alive

**HTTP Server Response:**

HTTP/1.0 200 OK

Content-Type: text/html

<html>

<body>

<span style="color:#0000A0">

<h1>Embedded Web Server</h1>

<h3>ATMEGA32 and ENC28J60</h3>

```
<p><form method="POST">

<strong>Temp: <input type="text" size=2 value="26"> <sup>O</sup>C

<p><input type="radio" name="radio" value="0" >Blinking LED

<br><input type="radio" name="radio" value="1" checked>Scanning LED

</strong><p>

<input type="submit">

</form></span>

</body>

</html>
```

The client then will translate this received HTML text and display the information on the browser screen such as room's temperature and the output LED status. By submitting different LED setting from the browser (POST request) to the HTTP server, now we could easily give the needed instruction to the AVR ATMEGA32 microcontroller that also functioned as the embedded web server.

To implement the concept of social network for smart devices, dynamic webpage development is essential. For this, server side programming is carried out using PHP language. PHP is a widely-used general-purpose scripting language that is especially suited for Web development and can be embedded into HTML. PHP is known as a server-sided language. That's because the PHP doesn't get executed on your computer, but on the computer you requested the page from. The results are then handed over to you, and displayed in your browser. To test the PHP scripts a server is required [26][27]. For the purpose of the project, Wampserver; a windows web development environment, an open source project is used so that the PC acts as a server [32].

The project is tested with different versions of browsers running on different operating systems. All browsers supporting HTTP V1.0 are able to communicate with the Microcontroller. And through the interface of the web page devices were controlled.

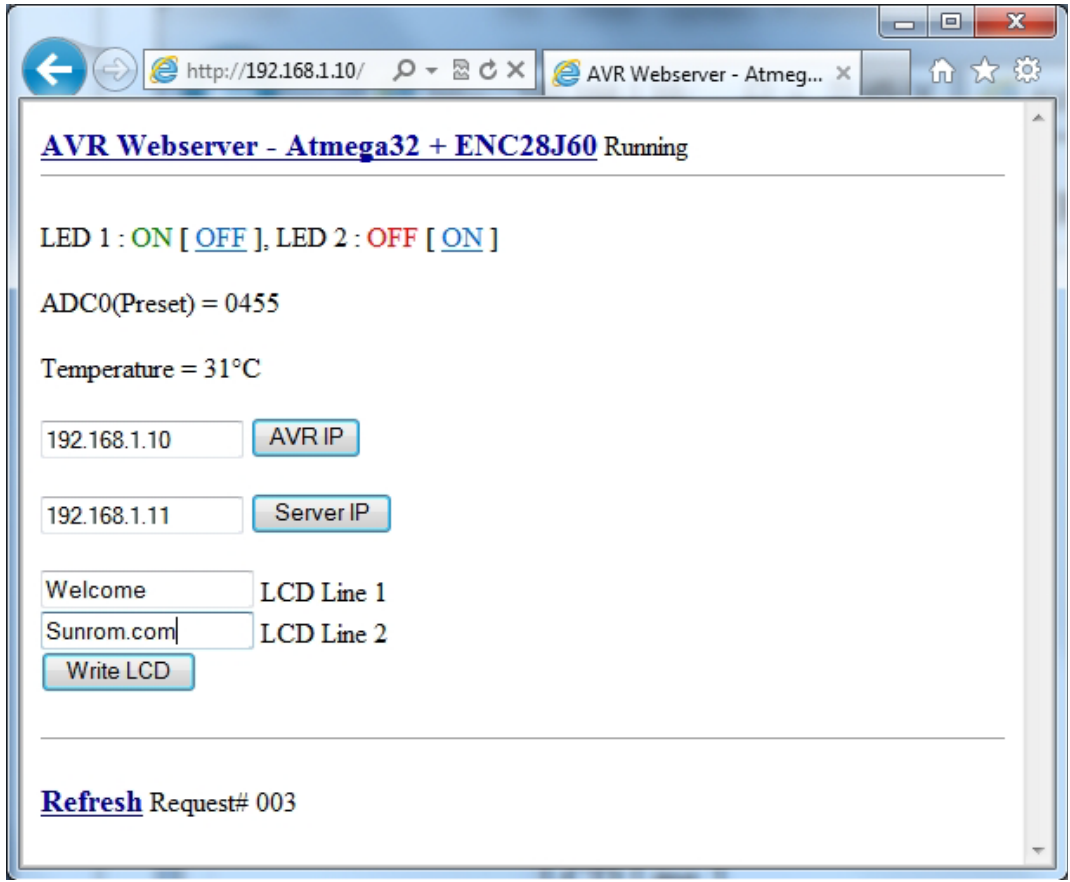

Figure 8. Final Webpage displaying status of the devices

## 6. CONCLUSIONS

The project on "Social Network for Smart Devices using Embedded Ethernet" has been developed which hosts a small web page. The web page is stored onto the ATmega32 Flash memory. Through the interface of the web page a user is able to control the devices remotely. A user can also monitor the devices remotely by embedding each device to be controlled or monitored with the AVR web server.In order to transmit the data from an existing device with SPI interface to network, an embedded Ethernet monitor and controlling system based on web browser is designed. This design can be used widely in remote data acquisition and control system in industry [3][12]. These Embedded Ethernet modules are having the capacity to perform as a true Ethernet device. It is possible to interface different kind of sensors with these modules and make various applications. So it can monitor embedded system operation state through Internet, achieving network monitoring purposes [14]. The AVR Web Server system adopts the high performance Ethernet controller, the system communication and debugging are fast, reliable and real-time; In addition, it can be also applied in on-line monitoring, remote fault diagnosis system [9]. The Project can be further developed to provide interfacing with SD/MMC cards or an IDE hard disk, so that a user can also upload files onto this web server through the web page interface. In a similar way the Microcontroller can be interfaced with a Camera and can be used for Remote Surveillance through web page [23].

# APPENDIX

### 1. Homepage (homepage.html)

\<title>AVR Webserver - Atmega32 + ENC28J60\</title>

\<a href="http://www.avrportal.com/" target="_blank">

\<b>\<font color="#000099" size="+1">AVR Webserver - Atmega32 + ENC28J60\</font>\</b>\</a>

Running\<br>\<hr width="100%" size="1">\<br>

LED 1 : \<font color=red>OFF\</font> [ \<a href="./?l1=1">ON\</a> ],

LED 2 : \<font color=red>OFF\</font> [ \<a href="./?l2=1">ON\</a> ]\<br>

\<br>ADC0(Preset) = 0482\<br>

\<br>Temperature = 27°C\<br>\<br>

\<form action="./?" method="get">\<input name="aip" type="text" size="15" maxlength="15" value="192.168.1.10">

 \<input type="submit" value="AVR IP">\</form>

 \<form action="./?" method="get">\<input name="sip" type="text" size="15" maxlength="15" value="192.168.1.11">

 \<input type="submit" value="Server IP">\</form>\<form action="./?" method="get">

 \<input name="lcd1" type="text" size="16" maxlength="16"> LCD Line 1\<br>

 \<input name="lcd2" type="text" size="16" maxlength="16"> LCD Line 2\<br>

 \<input type="submit" value="Write LCD">\</form>\<hr width="100%" size="1">\<br>\<a href="./">

 \<b>\<font color="#000099" size="+1">Refresh\</font>\</b>\</a> Request# 003

**2. PHP Script (index.php) [31]:**

```
<title>AVR Ethernet Logger</title>

<?

$filename = "./data.txt";

$handle = fopen ($filename, "r");

$contents = fread ($handle, filesize ($filename));

fclose ($handle);

//echo $contents;

$all = explode ( ',', $contents );

echo '<table><tr><td align="center"><b>Date & time<b></td><td
align="center"><b>Temparature</b></td></tr>';

for ($i = 0; $i < (count($all) - 1); $i++)

{

        $date_temp = explode( "|",$all[ $i ]  );

        echo '<tr><td align="center">'.$date_temp[0].'</td><td align="center">'.$date_temp[1].'
°C</td></tr>';

}

echo '</table>';

?>
```

# REFERENCES

[1] V L Varaprasad Nagula, M. Venkateswara Rao, T. Raghavendra Vishnu, "Embedded Ethernet Monitor and Controlling using Web Browser," International journal of Engineering Science & Advanced Technology(IJESAT), Vol 2, Special Issue 6, 1-5

[2] Mohammed Rahmatullah, "Web page analyser for protocols on ENC28J60 based AVR microcontroller," Int. J. Computer Technology & Applications, Vol 3 (3), 916-924

[3] Tao Lin ,Hai Zhao ,Jiyong Wang ,Guangjie Han and Jindong Wang ,"An Embedded Web Server for Equipment," School of Information Science & Engineering, Northeastern University, Shenyang, Liaoning, China.

[4] Feng Xia, Laurence T. Yang, Lizhe Wang, Alexey Vinel, "Internet of Things," International Journal of Communication Systems, Special Issue: Internet of Things, Volume 25, Issue 9, pages 1101–1102

[5] Zhao Hai, "Embedded Internet – an information technology revolution of 21st century," Beijing: Tsinghua University Press, 2002, pp. 198–225

[6] Luigi Atzori, Antonio Iera & Giacomo Morabito, "Computer Networks: The International Journal of Computer and Telecommunications Networking," Volume 54 Issue 15, October, 2010 Pages 2787-2805

[7] V.Sridhar, "Embedded Web Server," International Journal of Science, Engineering and Technology Research (IJSETR) Volume 1, Issue 3, September 2012

[8] Zhan Mei-qiong & Ji-Chang-peng, "Research and Implementation of Embedded Web Server," International Conference on MultiMedia and Information Technology, 2008, Pgs 123-125

[9] Indu Hariyale & V.A.Gulhane "Development of an Embedded Web Server System for Controlling and Monitoring of Remote Devices," 2nd National Conference on Information and Communication Technology, Number 1 - Article 3, 2011

[10] Srinivas Raja, G. Srinivas Babu, "Design of Web based Remote Embedded Monitoring system" International Journal of Technology and Engineering system(IJTES),Jan-March 2011-Vol.2,No.2.

[11] Fang Hongping, Fang KangLing, "The Design of Remote Embedded Monitoring System based on Internet" International Conference on Measuring Technology and Mechatronics Automation, 2010.

[12] Young-tao ZHOU, Xiao-hu CHEN,XU-ping WANG ,Chun- jiang YAO, "Design of equipment Remote Monitoring System Based on Embedded Web", International conference on Embedded software and Symposium(ICESS2008), 2008.

[13] Liu Yang, Linying Jiang , Kun Yue ,Heming Pang, "Design and Implementation of the Lab Remote Monitoring System Based on Embedded Web Technology" International Forum on Information Technology and Applications, 2010.

[14] Wang Xinxin, Chen Yun and Yan Ruzhong, "Implementation of the Web-based Mechanical and Electrical Equipment Remote Monitoring System", Computer Engineering, (31):231-233, 2005.

[15] Hakima Chaouchi, "The Internet of Things: Connecting Objects to the Web," Wiley

[16] Al Williams, "Embedded Internet Design," Tata McGraw –Hill, Edition, 2005

[17] Dreamtech Software Team, "Programming for Embedded Systems - Cracking the Code," WILEY-dreamtech India Pvt. Ltd., 2005

[18] Dr. K.V.K.K. Prasad, "Embedded/Real-Time Systems: Concepts, Design & Programming," Dreamtech Press, 2005

[19] Jan Axelson, "Embedded Ethernet and Internet Complete-Designing and Programming Small Devices for Networking"

[20] Jeremy Bentham, "TCP/IP-lean-webservers for embedded systems-2nd-edition," CMP Books, 2002

[21] Dan Eisenreich & Brian DeMuth, "Designing Embedded Internet Devices," Newnes, 2002

[22] Fred Eady, "Implementing 802.11 with Microcontrollers: Wireless Networking for Embedded Systems Designers," Newnes, 2005

[23] Jean-Philippe Vasseur, "Interconnecting Smart Objects with IP: The Next Internet," Morgan Kaufmann; 1 edition, 2010

[24] Cuno Pfister, "Getting Started with the Internet of Things: Connecting Sensors and Microcontrollers to the Cloud," O'Reilly Media; 1 edition, 2011

[25] Richard H. Barnett, Sarah Cox & Larry O'Cull "Embedded C Programming and the Atmel AVR," Delmar Cengage Learning; 2 edition, 2006

[26] Ian Llyod, "Build your own website the right way: Using HTML and CSS," SitePoint, 3rd Ed., 2011

[27] Kevin Yank, "Build your database driven website: Using PHP and MYSQL," Sitepoint, 4th Ed., 2009

[28] Atmel ATmega32 Datasheet, Atmel Corporation.
     www.atmel.com/dyn/resources/prod_documents/doc2503.pdf
[29] ENC28J60 Datasheet, Microchip Technology Inc.
     http://www.microchip.com/wwwproducts/Devices.aspx?dDocName=en022889
[30] Gary Desrosiers,99. Embedded Ethernet.http://www.embeddedethernet.com/
[31] Code Reference:AVR Web Server: www.avrportal.com
[32] Wampserver development environment: http://www.wampserver.com/
[33] AVR Studio:http://www.atmel.com/microsite/avr_studio_5/
[34] WinAVR:http://winavr.sourceforge.net

## AUTHORS

**Ketul Sheth**, Department of Electronics and Telecommunication Engineering, D.J. Sanghvi College of Engineering, Vile Parle(W), Mumbai, India. Currently pursuing 4th year Electronics and Telecommunication Engg.
Area of Interest: Embedded System Design and Computer Programming.

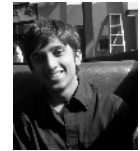

**Shreya Shah**, Department of Electronics and Telecommunication Engineering, D.J. Sanghvi College of Engineering, Vile Parle(W), Mumbai, India. Currently pursuing 4th year Electronics and Telecommunication Engg.
Area of Interest: Computer Networking, Network Programming.

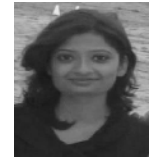

**Darshan Shah**, Department of Electronics and Telecommunication Engineering, D.J. Sanghvi College of Engineering, Vile Parle(W), Mumbai, India. Currently pursuing 4th year Electronics and Telecommunication Engg.
Area of Interest: Microcontroller based projects and Embedded Systems.

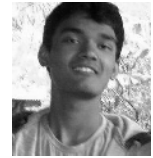

**Anuja Odhekar**, Assistant Professor, Department of Electronics and Telecommunication Engineering, D.J. Sanghvi College of Engineering, Vile Parle(W), Mumbai, India.
Area of Interest: Microwave and Embedded Systems

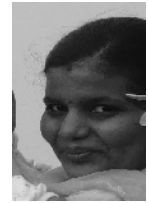